\documentclass[draft]{agujournal}
\journalname{JGR-Space Physics}

\begin{document}

%% ------------------------------------------------------------------------ %%
%  Title
%
% (A title should be specific, informative, and brief. Use
% abbreviations only if they are defined in the abstract. Titles that
% start with general keywords then specific terms are optimized in
% searches)
%
%% ------------------------------------------------------------------------ %%

% Example: \title{This is a test title}

\title{Explicit IMF $B_y$-effect maximizes at subauroral latitudes (Dedicated to the memory of Eigil Friis-Christensen)}

%% ------------------------------------------------------------------------ %%
%
%  AUTHORS AND AFFILIATIONS
%
%% ------------------------------------------------------------------------ %%

% Authors are individuals who have significantly contributed to the
% research and preparation of the article. Group authors are allowed, if
% each author in the group is separately identified in an appendix.)

% List authors by first name or initial followed by last name and
% separated by commas. Use \affil{} to number affiliations, and
% \thanks{} for author notes.
% Additional author notes should be indicated with \thanks{} (for
% example, for current addresses).

% Example: \authors{A. B. Author\affil{1}\thanks{Current address, Antartica}, B. C. Author\affil{2,3}, and D. E.
% Author\affil{3,4}\thanks{Also funded by Monsanto.}}

\authors{L. Holappa\affil{1,2,3}, N. Gopalswamy\affil{2} and K. Mursula\affil{1}}

\affiliation{1}{ReSoLVE Centre of Excellence, Space Physics Research Unit, University of Oulu, Oulu, Finland}
\affiliation{2}{NASA Goddard Space Flight Center, Greenbelt, MD, USA}
\affiliation{3}{Catholic University of America, Washington, DC, USA}
% \affiliation{4}{Fourth Affiliation}

%\affiliation{=number=}{=Affiliation Address=}
%(repeat as many times as is necessary)

%% Corresponding Author:
% Corresponding author mailing address and e-mail address:

% (include name and email addresses of the corresponding author.  More
% than one corresponding author is allowed in this LaTeX file and for
% publication; but only one corresponding author is allowed in our
% editorial system.)

% Example: \correspondingauthor{First and Last Name}{email@address.edu}

\correspondingauthor{Lauri Holappa}{lauri.holappa@oulu.fi}

%% Keypoints, final entry on title page.

% Example:
% \begin{keypoints}
% \item	List up to three key points (at least one is required)
% \item	Key Points summarize the main points and conclusions of the article
% \item	Each must be 100 characters or less with no special characters or punctuation
% \end{keypoints}

%  List up to three key points (at least one is required)
%  Key Points summarize the main points and conclusions of the article
%  Each must be 100 characters or less with no special characters or punctuation

\begin{keypoints}
\item IMF $B_y$-component is an explicit driver of geomagnetic activity, with the largest effect at subauroral latitudes
\item $B_y$-effect increases for strong solar wind driving and for winter conditions
\item Maximum $B_y$-effect is 20\% for all solar wind and 40\% for CMEs.

\end{keypoints}

%\item Magnetic clouds and their sheath regions are effective in driving strong geomagnetic activity at subauroral latitudes

%% ------------------------------------------------------------------------ %%
%
%  ABSTRACT
%
% A good abstract will begin with a short description of the problem
% being addressed, briefly describe the new data or analyses, then
% briefly states the main conclusion(s) and how they are supported and
% uncertainties.
%% ------------------------------------------------------------------------ %%

%% \begin{abstract} starts the second page

\begin{abstract}

The most important parameter in the coupling between solar wind and geomagnetic activity is the $B_z$-component of the interplanetary magnetic field (IMF).
However, recent studies have shown that IMF $B_y$ is an additional, independent driver of geomagnetic activity. 
We use here local geomagnetic indices from a large network of magnetic stations to study how IMF $B_y$ affects geomagnetic activity at different latitudes for all solar wind and, separately during coronal mass ejections (CMEs).
We show that geomagnetic activity, for all solar wind, is 20\% stronger for $B_y>0$ than for $B_y<0$ at subauroral latitudes of about $60^{\circ}$ corrected geomagnetic (CGM) latitude. 
During CMEs, the $B_y$-effect is larger, about 40\%, at slightly lower latitudes of about $57^{\circ}$ (CGM) latitude. 
These results highlight the importance of the IMF $B_y$-component for space weather at different latitudes and must be taken into account in space weather modeling.

%Thus, IMF $B_y$ is important for understanding and predicting space weather effects at auroral and subauroral latitudes.

\end{abstract}

%% ------------------------------------------------------------------------ %%
%
%  TEXT
%
%% ------------------------------------------------------------------------ %%

%%% Suggested section heads:
% \section{Introduction}
%
% The main text should start with an introduction. Except for short
% manuscripts (such as comments and replies), the text should be divided
% into sections, each with its own heading.

% Headings should be sentence fragments and do not begin with a
% lowercase letter or number. Examples of good headings are:

% \section{Materials and Methods}
% Here is text on Materials and Methods.
%
% \subsection{A descriptive heading about methods}
% More about Methods.
%
% \section{Data} (Or section title might be a descriptive heading about data)
%
% \section{Results} (Or section title might be a descriptive heading about the
% results)
%
% \section{Conclusions}

%Text here ===>>>

\section{Introduction}

% CME, MC and sheath

% global vs. local indices 

% By-effect 
Geomagnetic activity, the short-term variability of the Earth's magnetic field, is caused by the interaction of solar wind and the Earth's magnetic field.
The strongest magnetic disturbances on ground are due to auroral electrojets that are  located at about $70^{\circ}$ of corrected geomagnetic (CGM) latitude during average solar wind conditions. 
Severe space weather effects occur especially during geomagnetic storms, when the auroral region expands to subauroral or even lower latitudes.
Extensive areas of infrastructure are then exposed to strong magnetic disturbances caused by the auroral electrojets, as for example in 1989, when a major blackout occurred in Quebec, Canada \citep{Bolduc_2002}.        

Detailed understanding of the relation between solar wind and geomagnetic activity is important for space weather research and effects. 
It is well known that the strongest levels of solar wind driving occur during the Earth-passage of coronal mass ejections (CMEs) \citep{Gosling_1991, Gopalswamy_2005, Borovsky_2006, Zhang_2007}.
Coronal mass ejections observed at 1 AU often exhibit a magnetic cloud (MC) structure   with several distinguishing features, including a smooth rotation of the magnetic field and a low plasma density and pressure \citep{Burlaga_1988, Zurbuchen_2006}.   
Magnetic clouds moving faster than the magnetosonic speed  generate shocks and turbulent sheath regions which typically have highly variable magnetic fields and a high plasma density \citep{Kilpua_2013} and are also strong drivers of geomagnetic activity \citep{Huttunen_2002, Yermolaev_2012}.  

%Differences between MCs and sheath regions also lead to different responses in geomagnetic activity. legend('R^{+/-}(all SW, B_z < -5')
%While sheath regions are effective in driving auroral activity, MCs are more effective in driving strong geomagnetic storms \citep{Huttunen_2002, Yermolaev_2012}.   

% However, the $B_y$-dependence is partly due to the Russell-McPherron (RMP) effect \citep{Russell_1973}, which leads to negative $B_z$ for $B_y>0$ in fall and for $B_y<0$ in spring. 

The most critical solar wind parameter for geomagnetic activity is the $B_z$-component (measured in GSM coordinate system) of the interplanetary magnetic field (IMF), controlling reconnection rate in the dayside magnetopause.  
Both analytic work \citep{Sonnerup_1974} and MHD simulations [\citeauthor{Fedder_1991}, \citeyear{Fedder_1991}; \citeauthor{Laitinen_2007}, \citeyear{Laitinen_2007}] have shown that also IMF $B_y$-component affects the reconnection rate.
The IMF dependence of geomagnetic activity is often approximated by different coupling functions, such as the Newell universal coupling function \citep{Newell_2007}
\begin{equation}
d\Phi_{MP}/dt = v^{4/3}B_T^{2/3}\sin^{8/3}(\theta/2),
\end{equation}
where $v$ is solar wind speed, $B_T = \sqrt{B_z^2 + B_y^2}$ and $\theta = \arctan(B_y/B_z)$ is the so-called IMF clock angle.
In the Newell function $d\Phi_{MP}/dt$ (and in all other common coupling functions) the effect of $B_y$ is symmetric, that is, changing the sign of $B_y$ does not change the value of $d\Phi_{MP}/dt$. 
However, recent studies by \citet{Friis-Christensen_2017} and \citet{Smith_2017} showed that the $AL$-index (measuring the westward auroral electrojet) is considerably stronger for $B_y>0$ than for $B_y<0$ in Northern Hemisphere (NH) winter. 
Note that this \textit{explicit} $B_y$-dependency is not due to the Russell-McPherron (RMP) effect \citep{Russell_1973}, which maximizes in April and October, leading to a more negative $B_z$ even around northern winter (summer) solstice for $B_y<0$ ($B_y>0$).   
\citet{Holappa_2018} quantified this explicit $B_y$-effect to the westward electrojet in both hemispheres by removing the influence of the RMP effect, and showed that the $AL$-index is about 40-50\% stronger for $B_y>0$ than for $B_y<0$ around NH winter solstice. 
Even when averaged over all seasons and all solar wind data, $AL$-index is still about 12\% stronger for $B_y>0$ than for $B_y<0$.
\citet{Holappa_2018} also showed that the $B_y$-effect works oppositely the Southern Hemisphere, where $B_y<0$ yields to higher geomagnetic activity in local winter.

The exact physical mechanism of the explicit $B_y$-effect is still unknown.
Radar observations have shown that IMF $B_y$ affects the shape of the ionospheric convection patterns \citep{Ruohoniemi_2005, Pettigrew_2010}.
Recently, \citet{Thomas_2018} showed that for a given value of solar wind convective electric field, the cross-polar cap potential (measuring the strength of ionospheric convection) is greater for $B_y>0$ than for $B_y<0$ in winter, which is consistent with the above results based on geomagnetic activity.
\citet{Holappa_2018} showed that the $B_y$-effect in NH maximizes at 5 UT, that is, when the Earth's dipole axis points towards midnight and the NH ionosphere is maximally in darkness. 
The combined UT/seasonal variation of the $B_y$-effect indicates that the $B_y$-effect is most effective under low ionospheric conductivity. 
This is further supported by the fact that the $B_y$-effect in the SH maximizes in local winter. 
However, the $B_y$-dependencies in SH and NH are opposite: geomagnetic activity in the SH is higher for $B_y<0$ than for $B_y>0$.
%This shows that the $B_y$-effect does not modulate geomagnetic activity uniformly over the globe but shows a strong spatial 

%To derive these numbers, \citet{Holappa_2018} removed the $B_y$-dependence due to the Russell-McPherron effect \citep{Russell_1973}, in which the $B_y$-component is not a driver \textit{per se}, but rather modulates the $B_z$-component.

%Thus, on average the $B_y$-effect to the $AL$-index is rather small. 
From space weather perspective, it is crucial to quantify the significance of the $B_y$-effect at different latitudes. This has not yet been done in previous studies, which are all based on global geomagnetic indices.
This paper studies how the $B_y$-effect modulates geomagnetic activity at different latitudes by using local geomagnetic indices from a large network of magnetic stations. 
We will focus on periods of strong CME-driven geomagnetic activity. 
This paper is organized as follows.
In Section 2 we introduce the database of CMEs and other solar wind data, as well as geomagnetic indices used in this paper. 
In Section 3 we study the latitudinal distribution of geomagnetic activity during CMEs.
In Section 4 we study the effect of $B_y$ to local geomagnetic activity at different latitudes. 
In Section 5 we study the seasonal variation of the $B_y$-effect.    
Finally, we give our conclusions in Section 6.

\section{Data}

In this paper we use hourly averages of solar wind and IMF parameters measured in the GSM coordinate system from the OMNI2 database (\texttt{http://omniweb.gsfc.nasa.gov/}). 
We also use a list of 164 CMEs (magnetic clouds and associated sheath regions) identified from solar wind measurements by the Wind satellite at 1 AU in 1995-2015 \citep{Gopalswamy_2015}.
The primary identification criteria for MCs are low proton temperature and/or low plasma beta and smooth rotation rotation of IMF. 
(For a more detailed discussion on CME observations, see \citeauthor{Gopalswamy_2015}, \citeyear{Gopalswamy_2015}).

We use local measurements of geomagnetic activity from 44 stations in 1995-2016. The list of stations and their coordinates are given in Table 1. 
For all these stations we calculate their $A_h$-indices \citep{Mursula_Martini_2007_B} measuring local geomagnetic activity. 
$A_h$ indices are analogous to local $K$/$A_k$-indices, measuring the range of variation of the local horizontal magnetic field in three-hour intervals after removing the regular diurnal variation due to the solar quiet (Sq) currents in the ionosphere.
\citet{Mursula_Martini_2007_B} showed that the local $A_h$-indices correlate very well with the local $K$/$A_k$-indices, which are known to be good proxies for local GIC amplitudes \citep{Viljanen_2006}. 
Thus, the $A_h$-indices provide a well-suited database for studying the significance and space weather impact of the $B_y$-effect at different latitudes.

%However, unlike in the $K$-index method, the ranges measured by the $A_h$ index are taken as such, without converting to a quasi-logarithmic scale (numbers from 0 to 9).

\begin{table}
\caption{Stations and their corrected geomagnetic (CGM) and geographic (GG) coordinates. Stations are ordered according to their CGM latitudes. }
\hspace*{-1cm}
\begin{tabular}[l]{ccrrrr | ccrrrr}
\hline
\small
\# & Code & CGMlat & CGMlong & GGlat & GGlong & \# & Code & CGMlat & CGMlong & GGlat & GGlong \\\hline
1& ABG &10.37& 146.54 & 18.64 & 72.87 & 23& WNG & 49.96 & 86.38 & 53.74 & 9.07  \\
2& MBO &19.90& 57.82 & 14.38 & -16.97 & 24& HLP & 50.76 & 94.87 & 54.61 & 18.82  \\ 
3& HON & 21.28 & -89.48 & 21.32 & -158.00 &25& NVS & 50.84 & 156.56 & 54.85 & 83.23  \\ 
4& KNY & 24.87 & -156.48 & 31.42 & 130.88 &26& ESK & 52.57 & 77.04 & 55.32 & -3.20  \\
5& SJG & 28.31 & 6.57 & 18.38 & -66.12 & 27& VIC & 53.67 & -62.36 & 48.52 & -123.42 \\
6& KAK & 29.47 & -147.48 & 36.23 & 140.18 &28& NEW & 54.72 & -54.80 & 48.27 & -117.12  \\ 
7& BMT & 34.75 & -170.52 & 40.30 & 116.20 &29& NUR & 57.04 & 102.00 & 60.51 & 24.66  \\
8& MMB & 37.28 & -143.77 & 43.91 & 144.19 &30& LER & 57.87 & 80.49 & 60.13 & -1.18  \\
9& TUC & 39.81 & -44.07 & 32.25 & -110.83  &31& SIT & 59.67 & -78.19 & 57.05 & -135.34 \\
10& NCK & 42.78 & 91.49 & 47.63 & 16.72  &32& MEA & 61.67 & -52.10 & 54.62 & -113.35  \\
11& PAG & 42.81 & 98.63 & 47.48 & 24.18  &33& SOD & 64.09 & 107.04 & 67.37 & 26.63  \\
12& HRB & 43.08 & 92.79 & 47.87 & 18.19  &34& LRV & 64.64 & 66.11 & 64.18 & -21.70  \\
13& CLF & 43.32 & 79.20 & 48.02 & 2.27  &35& ABK & 65.44 & 101.72 & 68.36 & 18.82  \\
14& FUR & 43.57 & 87.31 & 48.17 & 11.28  &36& FCC & 68.32 & -25.96 & 58.79 & -94.09  \\
15& BDV & 44.58 & 89.82 & 49.08 & 14.02  &37& YKC & 69.15 & -57.04 & 62.48 & -114.48  \\
16& MAB & 46.18 & 82.95 & 50.30 & 5.68  &38& BRW & 70.27 & -106.53 & 71.30 & -156.62  \\
17& IRT & 47.24 & 177.95 & 52.17 & 104.45 &39& BLC & 73.33 & -30.33 & 64.33 & -96.03  \\
18& HAD & 47.37 & 74.46 & 51.00 & -4.48  &40& HRN & 74.34 & 108.24 & 77.00 & 15.55  \\
19& BEL & 47.65 & 96.08 & 51.84 & 20.79  &41& GDH & 75.05 & 38.41 & 69.25 & -53.53  \\
20& NGK & 47.95 & 88.96 & 52.07 & 12.68  &42& CBB & 76.81 & -47.90 & 69.12 & -105.03  \\
21& FRD & 48.33 & -1.09 & 38.21 & -77.37  &43& RES & 82.76 & -35.89 & 74.69 & -94.89  \\
22& BOU & 48.66 & -38.58 & 40.13 & -105.23  &44& THL & 84.64 & 28.23 & 77.48 & -69.17  \\
\hline
\end{tabular}
\hspace*{-1cm}
\end{table}

\section{Latitudinal distribution of geomagnetic activity driven by CMEs}

\begin{figure}[h!]
\centering
% when using pdflatex, use pdf file:
\includegraphics[width=20pc]{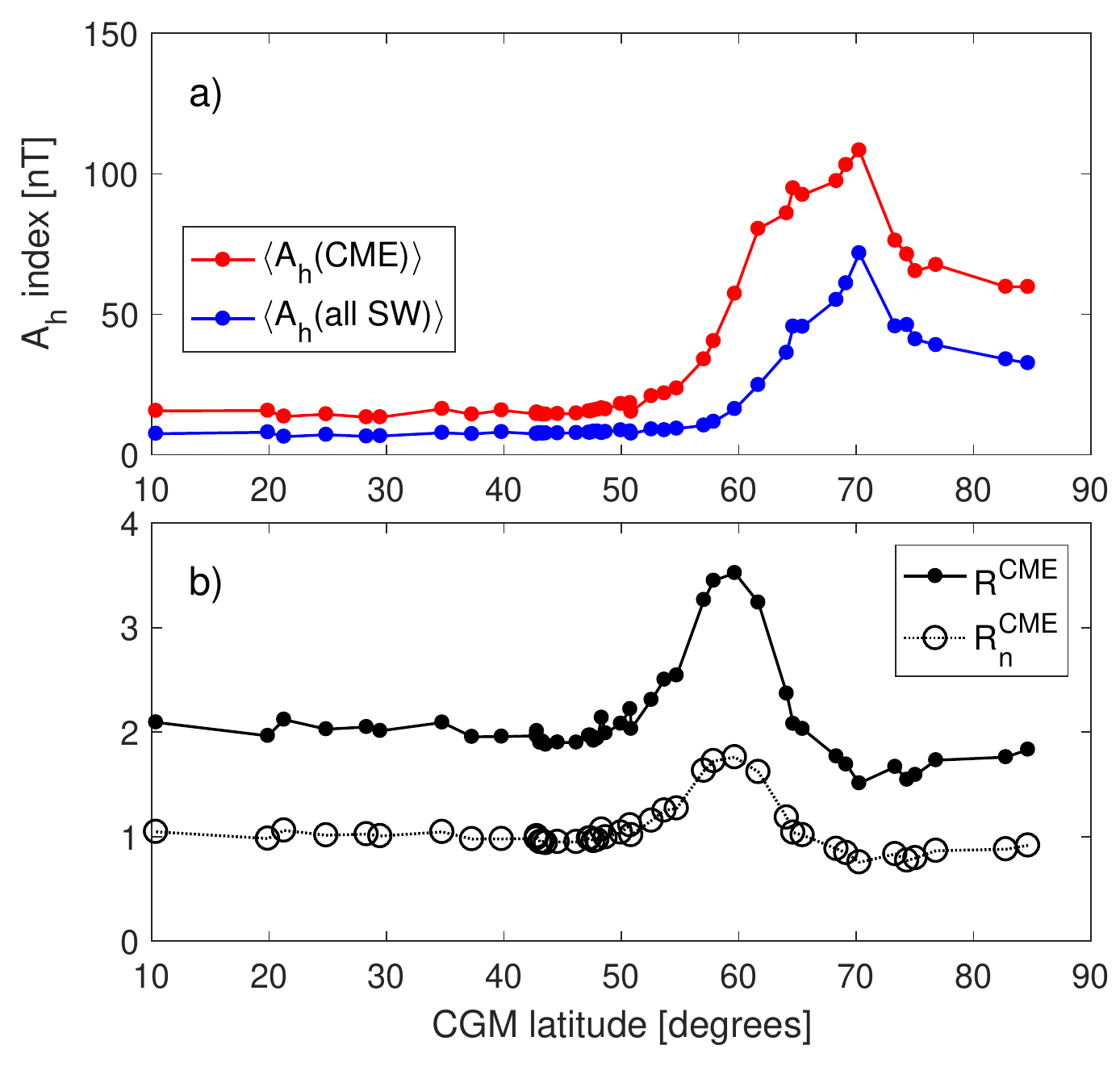}
\caption{a) Latitudinal distribution of the average $A_h$-indices during CMEs ($\langle A_h(CME)\rangle$) and all solar wind ($\langle A_h(all\, SW)\rangle)$. b) Latitudinal distributions of the ratio $R^{CME}$ = $\langle A_h(CME)\rangle/\langle A_h(all\, SW)\rangle$ and the normalized ratio $R^{CME}_n$ defined in Eq. (3). }
\label{lat_distr}
\end{figure}

Figure \ref{lat_distr}a shows the average values of $A_h$ indices during the 164 CMEs ($\langle A_h(CME) \rangle$) and for all solar wind data in 1995-2016 ($\langle A_h(all\, SW) \rangle$) as a function of the corrected geomagnetic latitude of the station.
Figure \ref{lat_distr}a verifies the well-known fact that geomagnetic activity is almost an order-of-magnitude stronger in the auroral region at about $65^{\circ}-70^{\circ}$ than at low latitudes. 
While $\langle A_h(all\, SW) \rangle$ shows a fairly sharp peak at $70^{\circ}$, $\langle A_h(CME) \rangle$ exhibits a clear broadening of the peak towards lower latitudes, with almost a plateau formed at about $64^{\circ}-68^{\circ}$. 

When averaged over all stations, $\langle A_h(CME) \rangle$ is 77\% greater than $\langle A_h(all\, SW) \rangle$.
However, there are latitudinal differences in the relative increase of geomagnetic activity.
This can be better seen in Figure \ref{lat_distr}b, which shows the ratio 
\begin{equation}
R^{CME} = \frac{\langle A_h(CME) \rangle}{\langle A_h(all\, SW) \rangle}
\end{equation} 
as a function of corrected geomagnetic latitude.
While the ratio $R^{CME}$ is almost a constant (about 2) at low and mid-latitudes, it reaches a peak of about 3.5 at subauroral latitudes (around $60^{\circ}$) and shows a minimum of about 1.5 in auroral latitudes (around $70^{\circ}$).
This is due to expansion of the auroral oval to lower latitudes due to strong driving by CMEs \citep{Borovsky_2006, Holappa_2014}. 
While geomagnetic activity increases at all latitudes during CMEs, the expansion of the auroral oval brings subauroral stations closer to the auroral electrojets, leading to a strong relative increase of activity at subauroral latitudes.

Expansion of the auroral oval during CMEs is further studied in Figure \ref{lat_distr}b, which shows the normalized ratio 
\begin{equation}
R^{CME}_n = R^{CME}\cdot \left(\frac{\langle d\Phi_{MP}/dt(CME) \rangle}{\langle d\Phi_{MP}/dt(all\, SW) \rangle}\right)^{-1} = \frac{\langle A_h(CME) \rangle}{\langle A_h(all\, SW) \rangle} \cdot \left(\frac{\langle d\Phi_{MP}/dt(CME) \rangle}{\langle d\Phi_{MP}/dt(all\, SW) \rangle}\right)^{-1}.  
\end{equation}
The ratio $R_n^{CME}$ is close to one at low and mid-latitudes.
Thus, the relative increase of the solar wind driving (quantified by the ratio $\langle d\Phi_{MP}/dt(CME) \rangle / \langle d\Phi_{MP}/dt(all\, SW) \rangle$) explains the relative increase of geomagnetic activity at these latitudes.
However, the ratio $R^{CME}_n$ peaks at subauroral latitudes (with a maximum of about 1.8) and is slightly below one at auroral latitudes (with a minimum of 0.8). 
Thus, the expansion of the auroral oval during CMEs leads to stronger (weaker) relative increase of geomagnetic activity at subauroral (auroral) latitudes.

%This is consistent with \citet{Huttunen_2002} who showed that sheaths are more likely to cause large values of the $K_p$-index than MCs, which is a weighed average of local $K$-indices in mid- and subauroral latitudes.

%he peak latitude of $R$ is found at about three degrees lower than the peak of $R(MC)$, indicating that sheath-driven substorms are stronger and possibly have onsets at lower latitudes than MC-driven substorms.  

%The stronger geoeffectiveness of sheaths than MCs 
%Both $R(MC)$ and $R(SH)$ minimize around $70^{\circ}$, where $\langle A_h(all) \rangle$ maximizes, indicating that 
%It is also noteworthy that the relative effectiveness of substorms drops  
%This is due to the fact that during strong solar wind driving of the magnetosphere by MCs and their sheath regions the polar cap expands and substorm onsets occur at lower latitudes \citep{Milan_2008, Holappa_2014}.

%\section{$B_y$-effect in global geomagnetic indices} 

\section{Effect of IMF $B_y$ at different latitudes} \label{by_sec}

%\subsection{Magnetic clouds and their sheath regions}

\begin{figure}[h!]
\centering
% when using pdflatex, use pdf file:
\includegraphics[width=20pc]{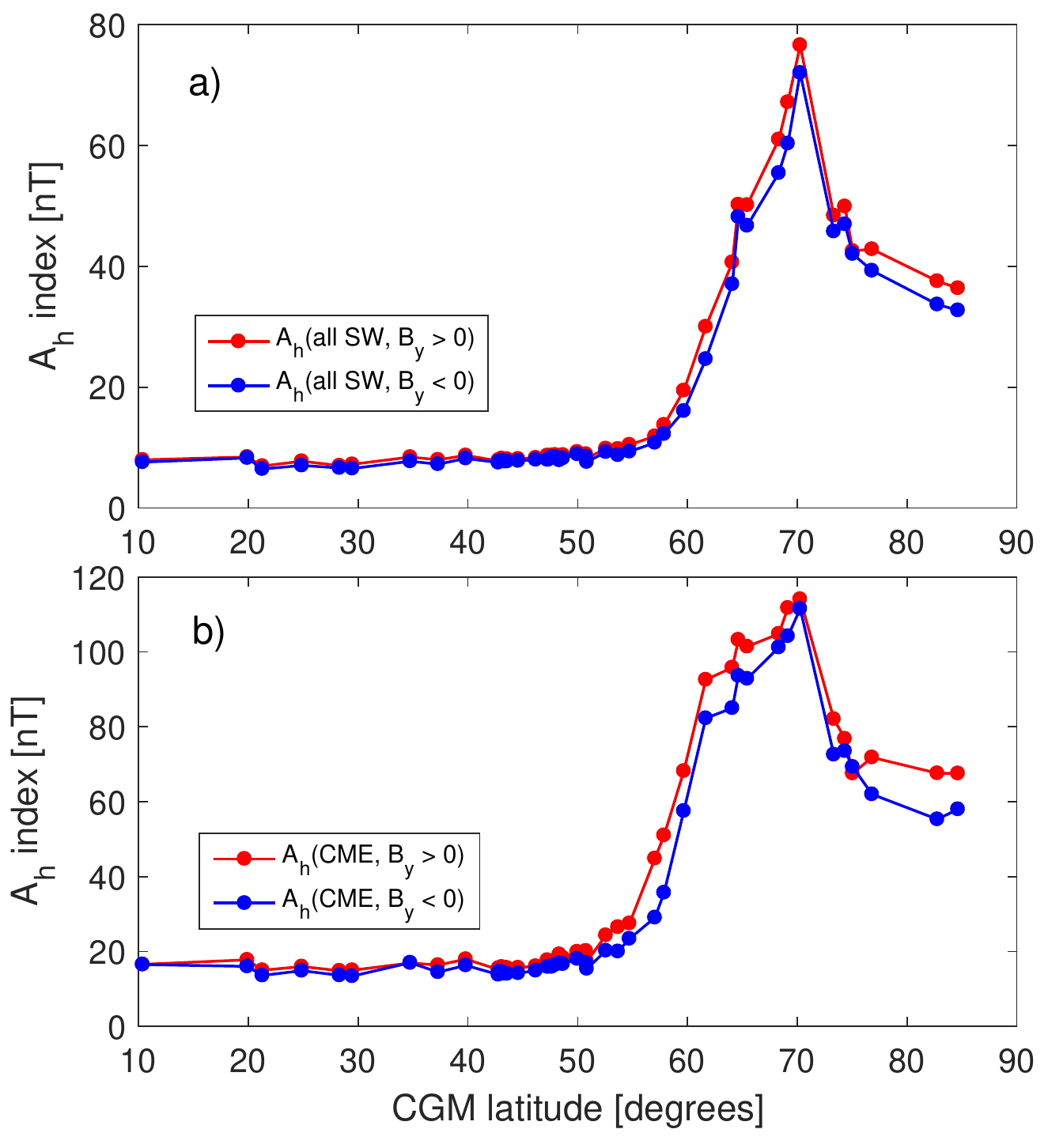}
\caption{Latitudinal distribution of $A_h$-indices for IMF $B_y>0$ and for $B<0$ during a) all solar wind b) CMEs.}
\label{lat_distr_by}
\end{figure}

%\added{The above analysis showed quantitatively how strong solar wind driving during CMEs leads to widening of the auroral oval and relatively strong enhancement of subauroral geomagnetic activity. 
%In this Section we study how IMF $B_y$ affects the latitudinal distribution of geomagnetic }
Figures \ref{lat_distr_by}a and \ref{lat_distr_by}b show the mean $A_h$ indices for $B_y>0$ and $B_y<0$ for all solar wind and during CMEs, respectively.
In Figure \ref{lat_distr_by}b the sign of $B_y$ has only a rather small effect. 
When averaging over all data and all stations, $\langle A_h\rangle$ indices are only 7.6\% stronger for $B_y>0$ than for $B_y<0$.
This is smaller than the 12\% $B_y$-effect to the $AL$-index found by \citet{Holappa_2018}. 
A smaller $B_y$-effect is understandable because the $AL$-index measures the strength of the westward electrojet (located mainly in midnight and dawn sectors), while the $A_h$-indices also include the effect of the eastward electrojet (afternoon sector), which is not affected by $B_y$ \citep{Holappa_2018}.
Interestingly, the $B_y$-effect is clearly stronger (12.4\%) for CMEs (Fig. \ref{lat_distr_by}b) than for all solar wind, when averaged over all stations. 

\begin{figure}[h]
\centering
% when using pdflatex, use pdf file:
\includegraphics[width=21pc]{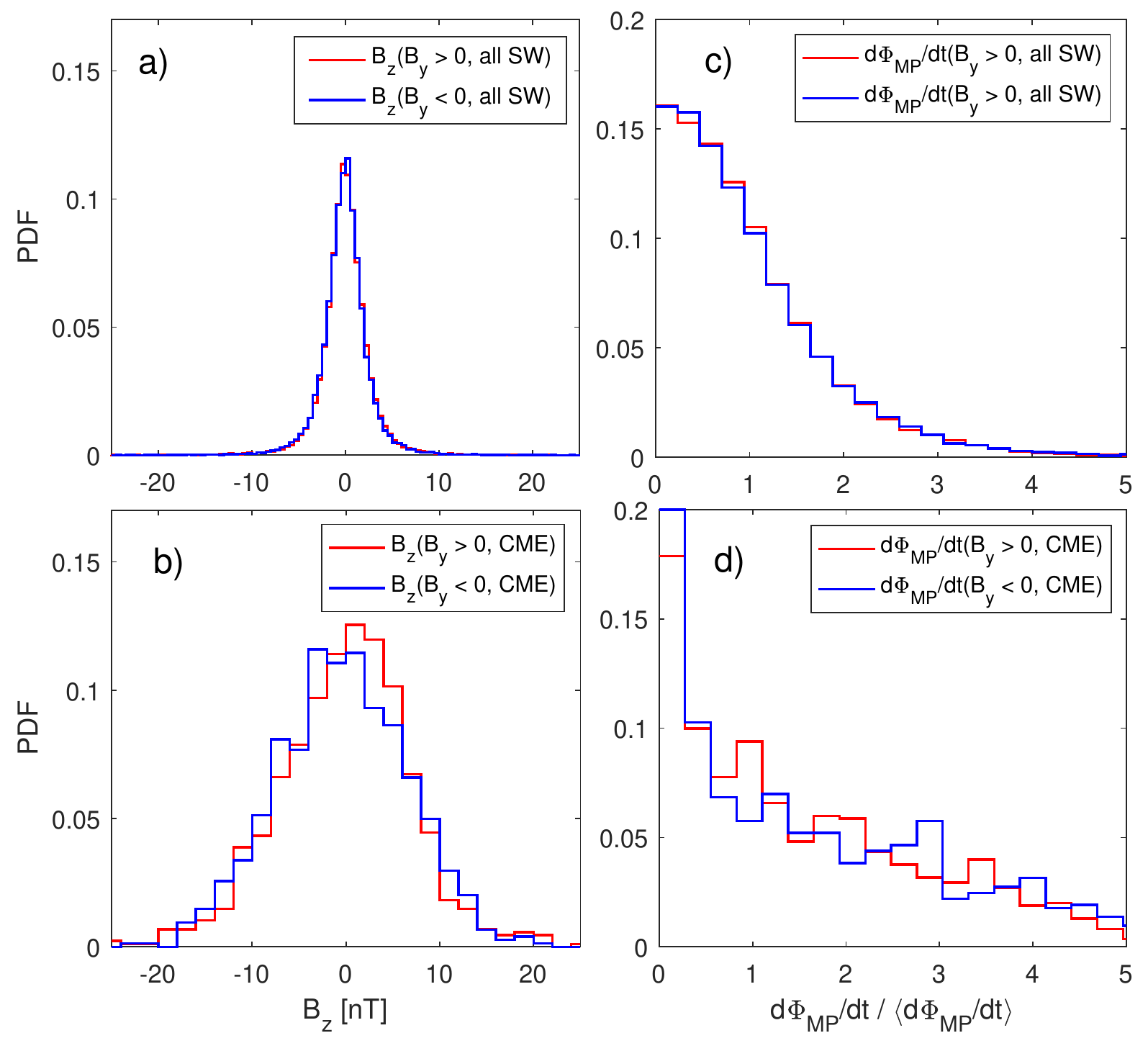}
\caption{a-b) Probability density functions of IMF $B_z$ in GSM coordinate system for CMEs and all solar wind. c-d) Probability density functions of the Newell universal couplung function $d\Phi_{MP}/dt$ in GSM coordinate system for CMEs and all solar wind. }
\label{Bz_distr}
\end{figure}

In order to rule out the possibility that the above $B_y$-effect is an artifact due to biased data selection, we calculate the probability distribution functions (PDF) of IMF $B_z$ and $d\Phi_{MP}/dt$  for $B_y>0$ and $B_y<0$.
Figures \ref{Bz_distr}a and \ref{Bz_distr}b show the PDFs of three-hour means of $B_z$ for all solar wind and for CME events, respectively.
Figure \ref{Bz_distr}a shows that, when all solar wind data are included in the statistics, the distribution of $B_z$ is virtually the same for $B_y>0$ and $B_y<0$. 
Moreover, Figure \ref{Bz_distr}b shows that even for the rather modest number of 164 CME events of our sample, the distributions of $B_z$ are almost equal for both signs of $B_y$. 
Figures \ref{Bz_distr}c and \ref{Bz_distr}d are similar to Figures \ref{Bz_distr}a and \ref{Bz_distr}b, but show the PDFs for three-hour means of $d\Phi_{MP}/dt$. 
Again, the distributions of $d\Phi_{MP}/dt$ are almost the same for $B_y>0$ and $B_y<0$.
Thus, there are no significant statistical differences in solar wind driving, which could explain the higher response in geomagnetic activity for $B_y>0$ than for $B_y<0$.

\begin{figure}[h!]
\centering
% when using pdflatex, use pdf file:
\includegraphics[width=21pc]{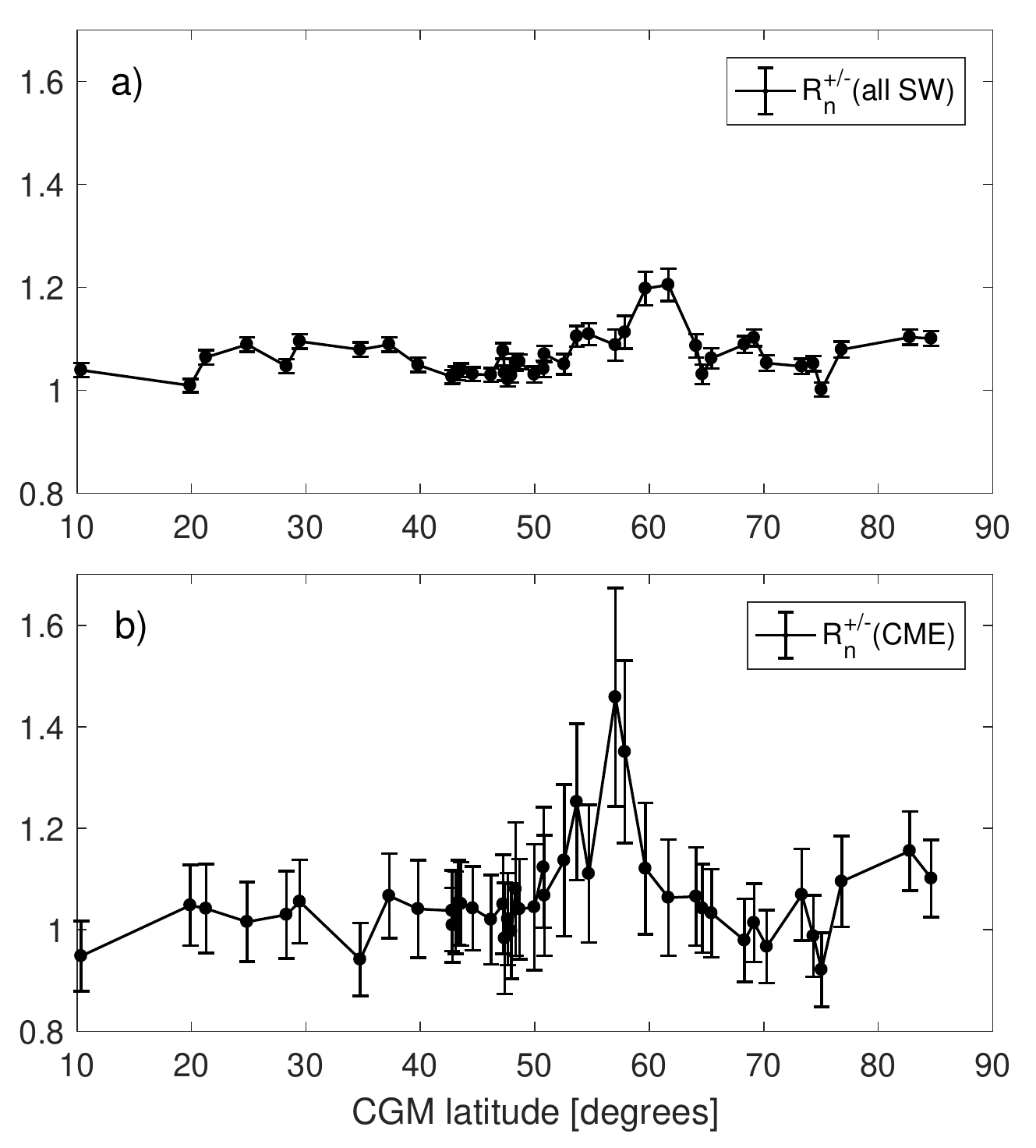}
\caption{Normalized ratios $R_n^{+/-}(all\, SW)$ (a) $R_n^{+/-}(CME)$ (b) defined in Equations (4-5). The vertical bars show the standard errors.}
\label{lat_distr_by_ratio}
\end{figure}

The relative size of the $B_y$-effect to geomagnetic activity at different latitudes is better seen in Figures \ref{lat_distr_by_ratio}a and \ref{lat_distr_by_ratio}b, which show the normalized ratios
\begin{equation}
R_n^{+/-}(all\, SW) = \frac{\langle A_h(all\, SW, B_y>0)\rangle}{\langle A_h(all\, SW, B_y<0)\rangle} \cdot \left(\frac{\langle d\Phi_{MP}/dt(all\, SW, B_y>0)\rangle}{\langle d\Phi_{MP}/dt(all\, SW, B_y<0)\rangle}\right)^{-1}
\end{equation} 
and 
\begin{equation}
R_n^{+/-}(CME) = \frac{\langle A_h(CME, B_y>0)\rangle}{\langle A_h(CME, B_y<0)\rangle} \cdot \left(\frac{\langle d\Phi_{MP}/dt(CME, B_y>0)\rangle}{\langle d\Phi_{MP}/dt(CME, B_y<0)\rangle}\right)^{-1},
\end{equation} 
respectively.
The most striking feature in Figure \ref{lat_distr_by_ratio}b is the high peak in $R_n^{+/-}(CME)$ ratio of about 1.5 peaking at subauroral latitudes of about $57^{\circ}$. 
The peak of $R_n^{+/-}(all\, SW)$ in Fig. \ref{lat_distr_by_ratio}a is also found at subauroral latitudes at about $60^{\circ}$, but the peak value (1.2) is considerably lower than for $R_n^{+/-}(CME)$. 
Both $R_n^{+/-}(CME)$ and $R_n^{+/-}(all\, SW)$ exhibit some irregularities in their latitudinal distributions, probably due to longitudinal dependence of the $B_y$-effect. 
(We leave the detailed analysis of longitudinal dependence out of this paper.)  
At latitudes below $45^{\circ}$ both $R_n^{+/-}(CME)$ and $R_n^{+/-}(all\, SW)$ are mostly slightly greater than one.   
Interestingly, both $R_n^{+/-}(CME)$ and $R_n^{+/-}(all\, SW)$ are only slightly greater than one at auroral latitudes (around $70^{\circ}$).
This indicates that the auroral electrojets are extended further to subauroral latitudes for $B_y>0$ than for $B_y<0$, decreasing the relative $B_y$-effect in auroral latitudes.
Figures \ref{lat_distr_by_ratio}a and \ref{lat_distr_by_ratio}b also include the standard errors of the two ratios, calculated by the formula \citep{Kendall_1994}
\begin{equation}
\sigma(R_n^{+/-}) \approx \frac{\langle A_h(B_y>0)\rangle}{\langle A_h(B_y<0)\rangle}\sqrt{\frac{\sigma(\langle A_h(B_y>0)\rangle)^2}{\langle A_h(B_y>0)\rangle^2} + \frac{\sigma(\langle A_h(B_y<0)\rangle)^2}{\langle A_h(B_y<0)\rangle^2} } \cdot \left(\frac{\langle d\Phi_{MP}/dt(B_y>0)\rangle}{\langle d\Phi_{MP}/dt(B_y<0)\rangle}\right)^{-1},
\end{equation}
where $\sigma(\cdot)$ denotes the standard error.
The relatively small sample size of 164 CMEs leads to considerably larger errors for CMEs than for all SW.

%The size of the $B_y$-effect at subauroral latitudes is surprisingly 
%Because sheath intervals constitute only a rather small fraction of 

\begin{figure}[h!]
\centering
% when using pdflatex, use pdf file:
\includegraphics[width=21pc]{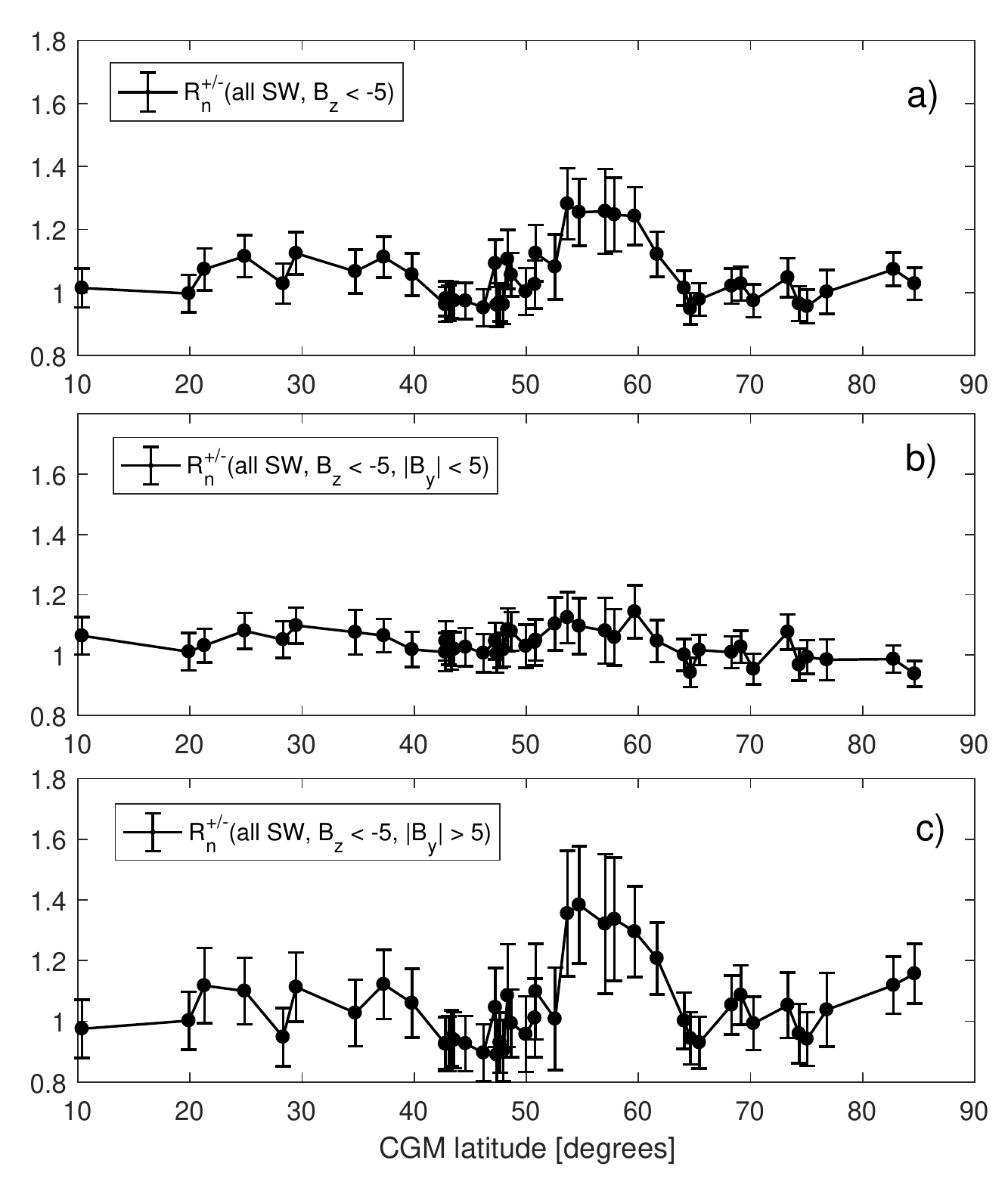}
\caption{a) Ratio $R_n^{+/-}(all\, SW, B_z < -5)$ defined in Eq. (7) b) ratio $R_n^{+/-}(all\, SW, B_z < -5, |B_y| < 5)$ c) ratio $R_n^{+/-}(all\, SW, B_z < -5, |B_y| > 5)$. Vertical bars denote the standard errors.}
\label{Ah_cme_like_distr}
\end{figure}

In order to further verify the robustness of the above results we have plotted in Figure \ref{Ah_cme_like_distr}a the ratio 
\begin{equation}
R_n^{+/-}(all\, SW,\, B_z < -5) = \frac{\langle A_h(B_z<-5, B_y>0)\rangle}{\langle A_h(B_z<-5, B_y<0)\rangle} \cdot \left(\frac{\langle d\Phi_{MP}/dt(all\, SW, B_y>0)\rangle}{\langle d\Phi_{MP}/dt(all\, SW, B_y<0)\rangle}\right)^{-1}
\end{equation}
 based on all 3-hour intervals solar wind data in 1995-2016 for which the three-hourly averaged $B_z < -5$ nT.
While this requirement does not exclusively identify CMEs from solar wind data, it ensures that solar wind driving is quite intense for a considerable time. 
Only 3.2\% of solar wind measurements meet this criterion (cf. Figure \ref{Bz_distr}a).
Because persistent strongly negative $B_z$ periods are commonly found within CMEs, but not, e.g, during CIRs/HSSs \citep{Tsurutani_1995, Yermolaev_2012}, the non-CME solar wind structures contribute to Figure \ref{Ah_cme_like_distr} quite little. 
Even though only a small fraction of all solar wind data is used in Figure \ref{Ah_cme_like_distr}, it is based on significantly larger statistics (1256 3-hr bins) than the results based on selected CME events (Figs. \ref{lat_distr_by}b, \ref{lat_distr_by_ratio}b and \ref{lat_distr_by_ratio}b).

Figure \ref{Ah_cme_like_distr}a shows that subauroral geomagnetic activity between $54^{\circ}-59^{\circ}$ is significantly greater for $B_y>0$ than for $B_y<0$ during strong solar wind driving. 
The peak of the ratio $R_n^{+/-}(all\, SW,\, B_z<-5)$ is about 1.3, in a close agreement with $R_n^{+/-}(CME)$ in Figure \ref{lat_distr_by_ratio}. 
However, the peak of $R_n^{+/-}(all\, SW, B_z<-5)$ extends to even lower latitudes than the peak of $R_n^{+/-}(CME)$. 
This indicates that the average level of geomagnetic activity is stronger under the condition $B_z<-5$nT than during CMEs, which include strongly negative but also strongly positive values of $B_z$.

In the above analysis we have only quantified the effect of the sign of $B_y$, without considering the amplitude $|B_y|$.
Figures \ref{Ah_cme_like_distr}b and \ref{Ah_cme_like_distr}c repeat the analysis of Figure \ref{Ah_cme_like_distr}a, imposing additional criterions: $|B_y|<5$ nT and $|B_y|>5$ nT, respectively.
The ratio $R_n^{+/-}(all\, SW,\, B_z<-5,\, |B_y|<5)$ in Fig. Fig. \ref{Ah_cme_like_distr}b is only slightly above one at most latitudes while the ratio $R_n^{+/-}(all\, SW,\, B_z<-5,\, |B_y|>5)$ reaches a maximum of about 1.4 between $54^{\circ}-59^{\circ}$. 
This indicates that the $B_y$-effect is only significant for rather strong values of $B_y$. 
Interestingly, Figures \ref{Ah_cme_like_distr}a and \ref{Ah_cme_like_distr}c show a local minimum at mid-latitudes at about $43^{\circ}-46^{\circ}$, where both ratios $R^{+/-}$ are close to one (within statistical error).

Figure \ref{Ah_cme_like_distr} highlights the importance of the $B_y$-effect for subauroral geomagnetic activity.   
Because the amplitude of IMF $B_y$ can be much larger than 5 nT, especially within CMEs, the $B_y$-effect can be even more important than in Figure \ref{Ah_cme_like_distr}c in extreme cases.

\section{Seasonal variation}

\begin{figure}[h!]
\centering
% when using pdflatex, use pdf file:
\includegraphics[width=21pc]{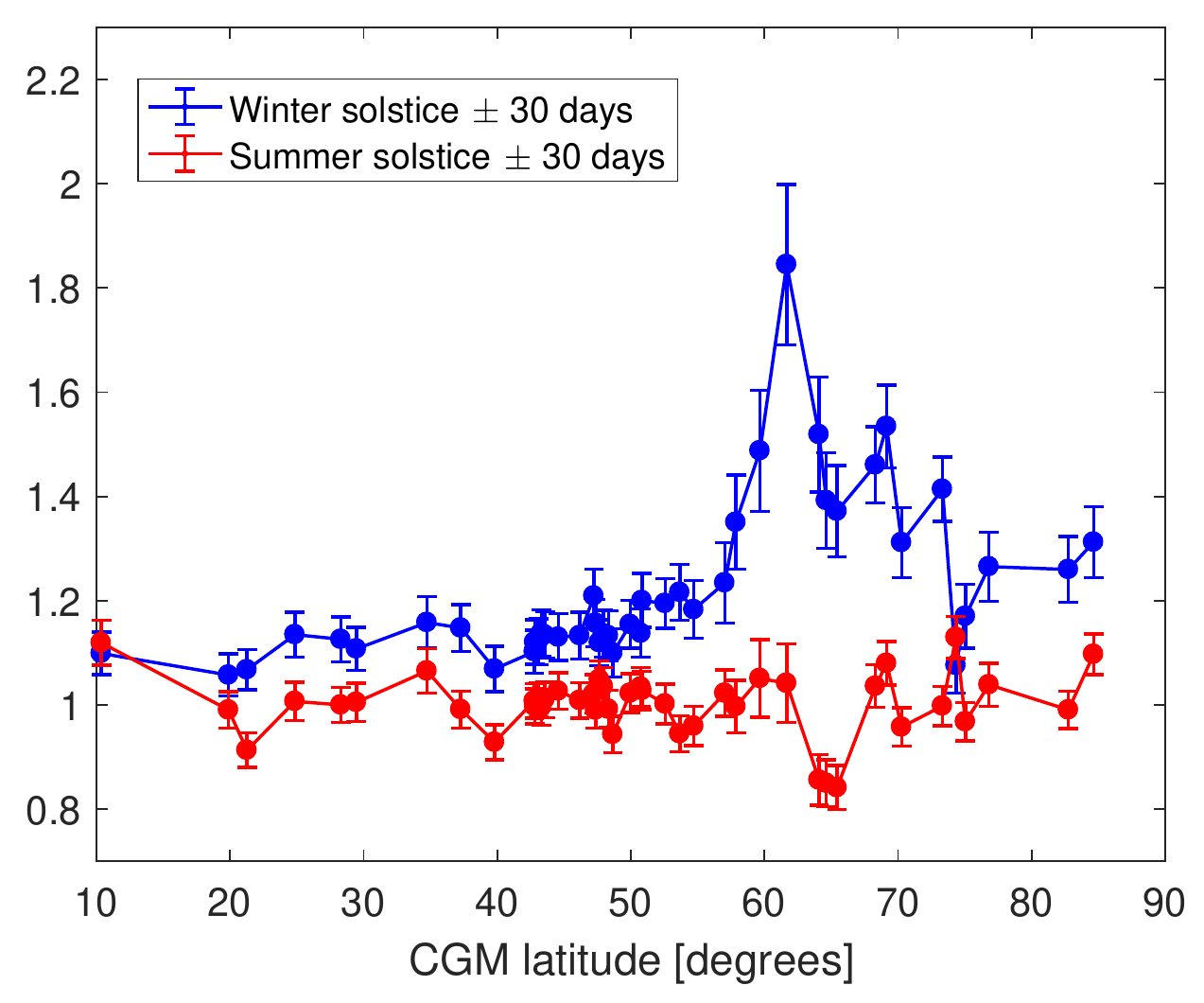}
\caption{Ratio $R_n^{+/-}(all\, SW,\, -3 < B_z < 0)$ calculated for winter and summer ($\pm 30$ days around winter and summer solstices, respectively)}
\label{seasonal}
\end{figure}

In the above analysis we have studied the $B_y$-effect by averaging over all seasons. However, as earlier studies \citep{Friis-Christensen_2017, Smith_2017, Holappa_2018} have shown, the $B_y$-effect is seasonally varying, maximizing in winter.
Figure \ref{seasonal} shows the ratio $R_n^{+/-}(all\, SW,\, -3< B_z < 0)$ separately for winter and summer ($\pm 30$ days around winter and summer solstices, respectively).
To have sufficient statistics, we have selected only periods of modest solar wind driving: -3 nT $<B_z<$ 0.
Figure \ref{seasonal} shows that subauroral geomagnetic activity at about $57^{\circ}-61^{\circ}$ is stronger for $B_y>0$ than for $B_y<0$ by a factor of 1.4-1.9 in winter. 
The peak of $R_n^{+/-}(all\, SW,\, -3 < B_z < 0)$ is at higher latitude ($61^{\circ}$) than in Figures \ref{Ah_cme_like_distr}a and \ref{Ah_cme_like_distr}b.
Note that for auroral latitudes, the ratio $R_n^{+/-}$ of Figure \ref{seasonal} gives roughly the same value of about 1.4 as earlier when using the auroral $AL$-index \citep{Holappa_2018}.
Figure \ref{seasonal} also shows that the maximum $B_y$-effect is not at the auroral latitudes. 
In summer the ratio $R_n^{+/-}(all\, SW,\, -3 < B_z < 0)$ is slightly below one at most latitudes, and it reaches a minimum of about 0.8 around $65^{\circ}$. 
This is in agreement with \citep{Holappa_2018} who found that the $AL$-index (measured between $60^{\circ}$ and $70^{\circ}$) is about 20\% weaker for $B_y>0$ than for $B_y<0$ around summer solstice.  

%While Figure \ref{seasonal} does not quantify the $B_y$-effect for strong solar wind driving in a similar way as Figure \ref{lat_distr_by_ratio} or \ref{Ah_cme_like_distr}, it shows that the $B_y$-effect works mainly in winter. 
%Because the $B_y$-effect during strong solar wind driving is as strong as 40\% when averaged over all seasons (Figs. \ref{lat_distr_by_ratio} and \ref{Ah_cme_like_distr}), it has to be even stronger during winter. 

\section{Discussion and conclusions}

In this paper we have studied the latitudinal distribution of the recently found explicit IMF $B_y$-dependence of geomagnetic activity for all solar wind and, separately, during coronal mass ejections. 
We find that the IMF $B_y$-component modulates geomagnetic activity for all solar wind and even more during CMEs, especially at subauroral latitudes.
During CMEs the $B_y$-effect maximizes at $59^{\circ}$ of corrected geomagnetic latitude, where local geomagnetic activity is about 40\% stronger for $B_y>0$ than for $B_y<0$.
The $B_y$-effect is relatively much stronger at subauroral latitudes than auroral latitudes, where it is only about 10\%.
This indicates that the auroral electrojets are latitudinally more extensive for $B_y>0$ than for $B_y<0$.

We also showed that a similar (about 30\%) $B_y$-effect at subauroral latitudes is observed for periods when the 3-hr average of IMF $B_z<-5$ nT. 
The size of the $B_y$-effect is even stronger (about 40\%) if, in addition to $B_z<-5$nT, we require $B_y$ to be large ($|B_y|>5$nT).

%We also find a fairly strong $B_y$-effect of about 20\% at low latitudes between 10^{\circ}-35^{\circ}.

The physical mechanism of the explicit $B_y$-dependence is not yet known.
\citet{Friis-Christensen_2017} showed that the $B_y$-effect mainly operates in the night sector and suggested that IMF $B_y$ modulates the strength of the substorm current wedge. 
This is supported by \citet{Holappa_2018} who showed that the $B_y$-effect modulates the $AL$-index (which is strongly affected by the substorm current wedge), but not the $AU$-index, which measures the eastward electrojet (not connected to the substorm current wedge).   
Under this assumption, our results suggest that the substorm current wedge extends to lower latitudes for $B_y>0$ than for $B_y<0$. 

Earlier studies have also shown that the $B_y$-effect exhibits a strong seasonal variation, maximizing in winter \citep{Smith_2017, Holappa_2018}.
In this paper we showed that the $B_y$-effect is important in winter at all latitudes.
We find at least a 10-20\% effect at all latitudes, with a maximum at subauroral latitudes, where $B_y>0$ yields nearly twice stronger geomagnetic activity than $B_y<0$.
The large winter $B_y$-effect supports the earlier finding that the $B_y$-effect maximizes when the ionosphere is maximally in darkness \citep{Holappa_2018}.
Thus, the the underlying mechanism of the $B_y$-effect is probably most efficient when the ionospheric conductivity is lowest. 
We also showed that during summer solstice the only significant $B_y$-effect is found at auroral latitudes of about $65^{\circ}$, where geomagnetic activity is about 20\% \textit{weaker} for $B_y>0$ than for $B_y<0$. 
% \added{The $B_y$-effect at subauroral latitudes is much stronger than the 40\% $B_y$-effect in the $AL$-index \citep{Holappa_2018}, measuring geomagnetic activity at auroral latitudes.} 

The results of this paper highlight the importance of the explicit IMF $B_y$-effect for understanding and predicting space weather effects at different latitudes, in particular during CMEs.

\acknowledgments

We acknowledge the financial support by the Academy of Finland to the ReSoLVE Centre of Excellence (project no. 272157).
The solar wind data were downloaded from the OMNI2 database (\texttt{http://omniweb.gsfc.nasa.gov/}).
Magnetometer data were downloaded from World Data Center, Edinburgh (\texttt{http://www.wdc.bgs.ac.uk/}).
\listofchanges
%%%


\begin{thebibliography}{27}
\providecommand{\natexlab}[1]{#1}
\expandafter\ifx\csname urlstyle\endcsname\relax
  \providecommand{\doi}[1]{doi:\discretionary{}{}{}#1}\else
  \providecommand{\doi}{doi:\discretionary{}{}{}\begingroup
  \urlstyle{rm}\Url}\fi

\bibitem[{\textit{Bolduc}(2002)}]{Bolduc_2002}
Bolduc, L. (2002), {GIC observations and studies in the Hydro-Qu{\'e}bec power
  system}, \textit{J. Atm.\ Sol.-Terr.\ Phys.}, \textit{64}(16), 1793--1802.

\bibitem[{\textit{{Borovsky} and {Denton}}(2006)}]{Borovsky_2006}
{Borovsky}, J.~E., and M.~H. {Denton} (2006), {Differences between CME-driven
  storms and CIR-driven storms}, \textit{J.\ Geophys.\ Res.}, \textit{111},
  A07S08, \doi{10.1029/2005JA011447}.

\bibitem[{\textit{{Burlaga}}(1988)}]{Burlaga_1988}
{Burlaga}, L.~F. (1988), {Magnetic clouds and force-free fields with constant
  alpha}, \textit{J.\ Geophys.\ Res.}, \textit{93}, 7217--7224,
  \doi{10.1029/JA093iA07p07217}.

\bibitem[{\textit{Fedder et~al.}(1991)\textit{Fedder, Mobarry, and
  Lyon}}]{Fedder_1991}
Fedder, J.~A., C.~M. Mobarry, and J.~G. Lyon (1991), Reconnection voltage as a
  function of imf clock angle, \textit{Geophys.\ Res.\ Lett.}, \textit{18}(6),
  1047--1050.

\bibitem[{\textit{Friis-Christensen et~al.}(2017)\textit{Friis-Christensen,
  Finlay, Hesse, and Laundal}}]{Friis-Christensen_2017}
Friis-Christensen, E., C.~C. Finlay, M.~Hesse, and K.~M. Laundal (2017),
  {Magnetic Field Perturbations from Currents in the Dark Polar Regions During
  Quiet Geomagnetic Conditions}, \textit{Space\ Sci.\ Rev.}, \textit{206}(1-4),
  281--297.

\bibitem[{\textit{Gopalswamy et~al.}(2005)\textit{Gopalswamy, Yashiro,
  Michalek, Xie, Lepping, and Howard}}]{Gopalswamy_2005}
Gopalswamy, N., S.~Yashiro, G.~Michalek, H.~Xie, R.~P. Lepping, and R.~A.
  Howard (2005), Solar source of the largest geomagnetic storm of cycle 23,
  \textit{Geophys.\ Res.\ Lett.}, \textit{32}(12).

\bibitem[{\textit{Gopalswamy et~al.}(2015)\textit{Gopalswamy, Yashiro, Xie,
  Akiyama, and M{\"a}kel{\"a}}}]{Gopalswamy_2015}
Gopalswamy, N., S.~Yashiro, H.~Xie, S.~Akiyama, and P.~M{\"a}kel{\"a} (2015),
  Properties and geoeffectiveness of magnetic clouds during solar cycles 23 and
  24, \textit{J.\ Geophys.\ Res.}, \textit{120}(11), 9221--9245.

\bibitem[{\textit{{Gosling} et~al.}(1991)\textit{{Gosling}, {McComas},
  {Phillips}, and {Bame}}}]{Gosling_1991}
{Gosling}, J.~T., D.~J. {McComas}, J.~L. {Phillips}, and S.~J. {Bame} (1991),
  {Geomagnetic activity associated with earth passage of interplanetary shock
  disturbances and coronal mass ejections}, \textit{J.\ Geophys.\ Res.},
  \textit{96}, 7831--7839, \doi{10.1029/91JA00316}.

\bibitem[{\textit{Holappa and Mursula}(2018)}]{Holappa_2018}
Holappa, L., and K.~Mursula (2018), {Explicit IMF $B_y$-dependence in
  high-latitude geomagnetic activity}, \textit{J.\ Geophys.\ Res.},
  \textit{123}, 4728--4740, \doi{10.1029/2018JA025517}.

\bibitem[{\textit{Holappa et~al.}(2014)\textit{Holappa, Mursula, Asikainen, and
  Richardson}}]{Holappa_2014}
Holappa, L., K.~Mursula, T.~Asikainen, and I.~G. Richardson (2014), Annual
  fractions of high-speed streams from principal component analysis of local
  geomagnetic activity, \textit{J.\ Geophys.\ Res.}, \textit{119}(6),
  4544--4555, \doi{10.1002/2014JA019958}.

\bibitem[{\textit{Huttunen et~al.}(2002)\textit{Huttunen, Koskinen, and
  Schwenn}}]{Huttunen_2002}
Huttunen, E.~K.~J., H.~E.~J. Koskinen, and R.~Schwenn (2002), Variability of
  magnetospheric storms driven by different solar wind perturbations,
  \textit{J.\ Geophys.\ Res.}, \textit{107}(A7), SMP 20--1--SMP 20--8,
  \doi{10.1029/2001JA900171}.

\bibitem[{\textit{Kendall et~al.}(1994)\textit{Kendall, Stuart, and
  Ord}}]{Kendall_1994}
Kendall, M., A.~Stuart, and J.~Ord (1994), \textit{Kendall's advanced theory of
  statistics. Vol. 1: Distribution theory}, Arnold, London.

\bibitem[{\textit{Kilpua et~al.}(2013)\textit{Kilpua, Hietala, Koskinen,
  Fontaine, and Turc}}]{Kilpua_2013}
Kilpua, E.~K.~J., H.~Hietala, H.~E.~J. Koskinen, D.~Fontaine, and L.~Turc
  (2013), Magnetic field and dynamic pressure ulf fluctuations in
  coronal-mass-ejection-driven sheath regions, \textit{Ann.\ Geophys.},
  \textit{31}(9), 1559--1567, \doi{10.5194/angeo-31-1559-2013}.

\bibitem[{\textit{Laitinen et~al.}(2007)\textit{Laitinen, Palmroth, Pulkkinen,
  Janhunen, and Koskinen}}]{Laitinen_2007}
Laitinen, T.~V., M.~Palmroth, T.~I. Pulkkinen, P.~Janhunen, and H.~E.~J.
  Koskinen (2007), Continuous reconnection line and pressure-dependent energy
  conversion on the magnetopause in a global mhd model, \textit{J.\ Geophys.\
  Res.}, \textit{112}(A11).

\bibitem[{\textit{{Mursula} and {Martini}}(2007)}]{Mursula_Martini_2007_B}
{Mursula}, K., and D.~{Martini} (2007), {New indices of geomagnetic activity at
  test: Comparing the correlation of the analogue ak index with the digital
  A$_{h}$ and IHV indices at the Sodankyl{\"a} station}, \textit{Adv.\ Space\
  Res.}, \textit{40}, 1105--1111, \doi{10.1016/j.asr.2007.06.067}.

\bibitem[{\textit{Newell et~al.}(2007)\textit{Newell, Sotirelis, Liou, Meng,
  and Rich}}]{Newell_2007}
Newell, P.~T., T.~Sotirelis, K.~Liou, C.-I. Meng, and F.~J. Rich (2007), A
  nearly universal solar wind-magnetosphere coupling function inferred from 10
  magnetospheric state variables, \textit{J.\ Geophys.\ Res.},
  \textit{112}(A1).

\bibitem[{\textit{Pettigrew et~al.}(2010)\textit{Pettigrew, Shepherd, and
  Ruohoniemi}}]{Pettigrew_2010}
Pettigrew, E.~D., S.~G. Shepherd, and J.~M. Ruohoniemi (2010), {Climatological
  patterns of high-latitude convection in the Northern and Southern
  hemispheres: Dipole tilt dependencies and interhemispheric comparisons},
  \textit{J.\ Geophys.\ Res.}, \textit{115}(A7), {A07305},
  \doi{10.1029/2009JA014956}, a07305.

\bibitem[{\textit{Ruohoniemi and Greenwald}(2005)}]{Ruohoniemi_2005}
Ruohoniemi, J.~M., and R.~A. Greenwald (2005), {Dependencies of high-latitude
  plasma convection: Consideration of interplanetary magnetic field, seasonal,
  and universal time factors in statistical patterns}, \textit{J.\ Geophys.\
  Res.}, \textit{110}(A9).

\bibitem[{\textit{{Russell} and {McPherron}}(1973)}]{Russell_1973}
{Russell}, C.~T., and R.~L. {McPherron} (1973), Semiannual variation of
  geomagnetic activity, \textit{J.\ Geophys.\ Res.}, \textit{78}(1), 92--108.

\bibitem[{\textit{Smith et~al.}(2017)\textit{Smith, Beggan, Macmillan, and
  Whaler}}]{Smith_2017}
Smith, A. R.~A., C.~D. Beggan, S.~Macmillan, and K.~A. Whaler (2017),
  {Climatology of the auroral electrojets derived from the along-track gradient
  of magnetic field intensity measured by POGO, Magsat, CHAMP, and Swarm},
  \textit{Space Weather}, \textit{15}(10), 1257--1269,
  \doi{10.1002/2017SW001675}, 2017SW001675.

\bibitem[{\textit{Sonnerup}(1974)}]{Sonnerup_1974}
Sonnerup, B.~U.~O. (1974), Magnetopause reconnection rate, \textit{J.\
  Geophys.\ Res.}, \textit{79}(10), 1546--1549, \doi{10.1029/JA079i010p01546}.

\bibitem[{\textit{Thomas and Shepherd}(2018)}]{Thomas_2018}
Thomas, E.~G., and S.~G. Shepherd (2018), {Statistical patterns of ionospheric
  convection derived from mid-latitude, high-latitude, and polar SuperDARN HF
  radar observations}, \textit{J.\ Geophys.\ Res.}, \textit{123}(4),
  3196--3216.

\bibitem[{\textit{{Tsurutani} et~al.}(1995)\textit{{Tsurutani}, {Gonzalez},
  {Gonzalez}, {Tang}, {Arballo}, and {Okada}}}]{Tsurutani_1995}
{Tsurutani}, B.~T., W.~D. {Gonzalez}, A.~L.~C. {Gonzalez}, F.~{Tang}, J.~K.
  {Arballo}, and M.~{Okada} (1995), {Interplanetary origin of geomagnetic
  activity in the declining phase of the solar cycle}, \textit{J.\ Geophys.\
  Res.}, \textit{100}, 21,717--21,734, \doi{10.1029/95JA01476}.

\bibitem[{\textit{Viljanen et~al.}(2006)\textit{Viljanen, Pulkkinen, Pirjola,
  Pajunp{\"a}{\"a}, Posio, and Koistinen}}]{Viljanen_2006}
Viljanen, A., A.~Pulkkinen, R.~Pirjola, K.~Pajunp{\"a}{\"a}, P.~Posio, and
  A.~Koistinen (2006), Recordings of geomagnetically induced currents and a
  nowcasting service of the finnish natural gas pipeline system, \textit{Space
  Weather}, \textit{4}(10).

\bibitem[{\textit{Yermolaev et~al.}(2012)\textit{Yermolaev, Nikolaeva, Lodkina,
  and Yermolaev}}]{Yermolaev_2012}
Yermolaev, Y.~I., N.~S. Nikolaeva, I.~G. Lodkina, and M.~Y. Yermolaev (2012),
  Geoeffectiveness and efficiency of cir, sheath, and icme in generation of
  magnetic storms, \textit{J.\ Geophys.\ Res.}, \textit{117}(A9).

\bibitem[{\textit{{Zhang} et~al.}(2007)\textit{{Zhang}, {Richardson}, {Webb},
  {Gopalswamy}, {Huttunen}, {Kasper}, {Nitta}, {Poomvises}, {Thompson}, {Wu},
  {Yashiro}, and {Zhukov}}}]{Zhang_2007}
{Zhang}, J., I.~G. {Richardson}, D.~F. {Webb}, N.~{Gopalswamy}, E.~{Huttunen},
  J.~C. {Kasper}, N.~V. {Nitta}, W.~{Poomvises}, B.~J. {Thompson}, C.-C. {Wu},
  S.~{Yashiro}, and A.~N. {Zhukov} (2007), {Solar and interplanetary sources of
  major geomagnetic storms ({Dst} $<$ -100 nT) during 1996-2005}, \textit{J.\
  Geophys.\ Res.}, \textit{112}, A10,102, \doi{10.1029/2007JA012321}.

\bibitem[{\textit{{Zurbuchen} and {Richardson}}(2006)}]{Zurbuchen_2006}
{Zurbuchen}, T.~H., and I.~G. {Richardson} (2006), {In-Situ Solar Wind and
  Magnetic Field Signatures of Interplanetary Coronal Mass Ejections},
  \textit{Space\ Sci.\ Rev.}, \textit{123}, 31--43,
  \doi{10.1007/s11214-006-9010-4}.

\end{thebibliography}
\end{document}